Housing Investment, Stock Market Participation and Household Portfolio choice:

Evidence from China's Urban Areas

Huirong Liu

Wuhan University

**Abstract**

This paper employs the survey data of CHFS (2013) to investigate the impact of housing investment on household stock market participation and portfolio choice. The results show that larger housing investment encourages the household participation in the stock market, but reduces the proportion of their stockholding. The above conclusion remains true even when the endogeneity problem is controlled with risk attitude classification, Heckman model test and subsample regression. This study shows that the growth in the housing market will not lead to stock market development because of lack of household financial literacy and the low expected yield on stock market.

*Key word:* Housing investment; Stock market participation; Portfolio choice

Housing Investment, Stock Market Participation and Household Portfolio choice: Evidence from China's Urban Areas

Since the turn of the century, China's housing market has thrived unprecedentedly, with housing property being integral to household assets due to the housing reform. While the housing market is booming, the stock market and other financial product markets in China witnessed a relative downturn. Take Shanghai Stock Exchange as an example, since the Shanghai composite index tumbled from the summit of 6124 in 2007, it has remained at a low level for 5 years. The weak stock market made many families lose their interest in stocks, and some even quit the stock market once for all.

The strong housing market and the weak stock market have sparked a fresh discussion. Some believe that the unprecedented development of the housing market has placed a cap on other financial markets. Does such crowding out play a part? Precisely, how does housing influence household stockholding?

From the perspective of household portfolio choice, conventional financial theory holds that rational economic men should diversify their investment across all risky assets with a certain proportion according to their different risk preferences (Markowitz, 1952; Tobin, 1958; Sharpe, 1964; Samuelson, 1969; Dow & da Costa Werlang, 1992). Hence, we must first answer two more practical questions to address the above questions. That is, how does housing investment affect household participation in the stock market? And how does housing investment affect the proportion of a household's stock investment?

Theoretically, the impact of housing investment on household's stock market participation and portfolio choice is uncertain: on the one hand, the rapid growth of house price after housing investment brings huge profit, which will stimulate families to invest in higher risky assets (Tobin, 1982). On the other, the temporary high-yield, low-risk nature of housing investment will attract the inflow of household funds, thus squeezing out the household's investment in stock assets and reducing their proportion of stockholding.

Based on model analysis, this paper employs the data from China Household Finance Survey (hereinafter referred to as CHFS) in 2013 to address the above two problems. Research results show that housing investment increases the probability of residents' participation in the stock market, but reduces the proportion of residents' stockholding. This conclusion still holds after the endogeneity problem is taken care of by risk attitude classification, Heckman model test and subsample regression. This result means that the "wealth effect" and "crowding out effect" both play a part in housing investment's impact on household stock investment.

Apart from the introduction, this paper is divided into five parts. The second part features a literature review where we give a brief introduction to the literature about the impact of housing investment on household financial decision-making at home and abroad. The third part is the research design, data introduction and variable selection. It specifically designs the research problems, introduces the data sources, and explains the data used in this paper and the variables used in the analysis. The forth part is empirical specifications, which explain how the models are established. The fifth part is the measurement results and discussion followed by a conclusion.

**Literature Review**

Since Calvet, Campbell, and Sodini (2009) proposed the concept of "household finance", scholars have been studying the household portfolio choice. The impact of housing investment on the household financial market participation and portfolio choice stands as a critical research topic.

As an asset class, properties are distinctive. On the one hand, the huge wealth effect brought by the appreciation of housing assets prompts families to hold more risky assets. Early theoretical literature has pointed out that, as people's relative risk aversion is a decreasing function of wealth, families will invest more in risky assets including stocks with the increase of wealth (Cohn, Lewellen, Lease, & Schlarbaum, 1975; Friend and Blume, 1975), while the property, as the most important household asset, has substantial wealth effect because it promotes the household's holding of risky assets (Tobin, 1982). Besides, in countries or regions with well-developed credit system, housing can be used as collateral to lay the foundation for families to invest in risky assets (Cardak & Wilkins, 2009).

But on the other hand, if the property investment temporarily emerges as high-yield and low-risk, while the stock investment temporarily appears to be the reverse, it will attract the inflow of household funds, thus squeezing out the household's stock investment and reducing the proportion of stockholding. X. Zhu and Y. Zhu (2014) analyzed that the residents are reluctant to participate in the stock market due to the unappealing risk element of the stock market, lack of information, and the transaction cost. Lee, Rosenthal, Veld, and Veld-Merkoulova (2015) found that the household's stock market expectation is significantly affected by the recent stock market condition, and the stock market expectation will decline sharply with the deterioration of the stock market environment. When the expectation for stock market goes bad, the household stock market investment will decrease accordingly (Vissing-Jorgensen, 2005; Hurd, van Rooij, & Winter, 2011). Zhang and Yin (2016) argued that the lack of financial literacy can significantly improve the probability of households' financial exclusion in China. Moreover, higher housing valuation may also hinder the household's risky assets holding. Fratantoni (1998) believes that owning a property brings price risk and committed expenditure risk to the household, which will push the household to reduce risky assets holding. Grossman and Laroque (1990) analyzed the impact of durable consumer goods such as housing on household portfolio choice. They pointed out that holding higher-value durable consumer goods will increase the liquidity risk of households, so higher property holding may lead to reduced holdings of risky assets. Flavin and Yamashita (2002) also believed that liquidity constraints caused by property purchase would hinder the household's investment in risky assets. Chetty and Szeidl (2007) proposed that committed consumption such as housing may increase the risk aversion of ordinary residents, contributing to a lowered proportion of risky assets holding.

From the empirical viewpoint, the results obtained with data from different countries may differ. Even when different data from the same country is used, the results may be different.

Vissing-Jorgensen (2005)'s research in the U.S. found that the income growth and the assets accumulation expand household financial ability to pay for stock investment cost and promote their participation in the stock market. Yamashita (2003) used the SCF data in 1989 to find that higher proportion of property in household assets is positively related to lower proportion of stock holding in household assets. Kullmann and Siegel's (2003) study in the U.S. found that housing investment reduces the household participation in stock market and the holding share of risky assets such as stocks. Cocco (2005) studied the Panel Survey of Income Dynamics (PSID) data in the U.S. and found that housing investment will crowd out stock investment, which is particularly prominent among young people and poor families. The Research on Australia by Cardak and Wilkins (2009) found that in countries and regions with better-established credit system, housing as collateral can create opportunity for households to invest in risky assets such as stocks. Chetty and Szeidl (2010) studied Survey of Income and Program Participation (SIPP) in the U.S., and found that housing mortgage would inhibit households' holding of risky assets. Fougere and Pouhes (2012) repeated Chetty and Szeidl (2010)'s study with French data, and still reached similar conclusions. Wu and Lu (2013), Wu and Qi (2007) study in China found that rising household housing investment brings reduced holding proportion of risky assets such as stocks, funds and foreign exchange in financial assets. However, Chen, Shi, and Quan (2015) used the CHFS (2011) data and found that a rising housing market will increase the value of household housing wealth, thus expanding residents' holding of stocks and other financial assets

In a word, although theretically, housing investment may both promote and inhibit household participating financial markets, the impact may differ by different contexts of various countries. In recent years, China's macro environment is relatively special given the general lacking of financial literacy among Chinese residents, continuing stagnant stock market and other financial product markets and meanwhile soaring housing prices. Therefore, this paper uses the data of CHFS in 2013 to explore the impact of Chinese residents' housing investment on their stock investment in recent years.

## Research Design and Data

**Research Design**

Because for Chinese residents, even though the proportion of housing in net assets stays the same, their attitude towards stock investment is different when the specific property value changes. Thus, the cost of housing investment is taken as an explanatory variable, rather than the proportion of the cost of housing investment in the total investment cost.

In this model, the housing investment cost is the explanatory variable, whether the household participates in the stock market is a response variable, with the other being the proportion of the stock investment cost to the total investment cost. Because the household investment behavior does not always occur at the same time, but in order to simplify the model, this paper does not consider the above factors. That is, the housing investment cost is the nominal expense of the housing investment expenditure at that time, and the stock investment cost is the nominal expense of the stock investment expenditure at that time. The total investment cost is the sum of the nominal expense of each investment, which is not converted into the present value based on the inflation rate.

This study also strips out the value-added part of housing investment and of the stock that cause the "wealth effect", and introduce two control variables: the added value of the housing wealth and the added value of the stock wealth. The value-added part of housing wealth is the difference between the total value of housing owned by the household and the cost of purchase. The value-added part of stock wealth is the difference between the current market value of the stock owned by the household and the cost of purchase.

**Data**

The data used in this paper is from the 2013 survey of China Household Finance Survey (CHFS), a national survey conducted by China Household Finance Survey and Research Center of Southwest University of Finance and Economics, which is conducted every other year. The sample covers a total of 29 provinces, 262 counties, 1048 communities and more than 27000 families across China. The survey includes the demographic characteristics of household members, household assets and liabilities, income status, investment concept and other aspects (Zhang & Yin, 2016).

The above information lays a solid foundation for this paper to study the participation decision and holding ratio of housing to household stock investment. Because only urban housing in China has a relatively clear transaction price, this paper only retains the urban samples and takes "housing" as the core explanatory variable. Likewise, this paper excludes the samples where the properties are not owned by household members. After clearing up the missing value and outliers, the final sample includes 13984 families.

**Descriptive Statistics**

The main response variables in this paper are "ifstock" and "stock_ratio". "Ifstock" indicates whether the household holds a stock account. If it holds, 1 is taken; if it does not, 0 is taken. "Stock_ratio" is the abbreviation of "the proportion of stock investment cost in the total cost of investment assets". The investment assets defined in this paper mainly include housing and financial assets. According to Yin, Song, and Wu (2014) and Yin, Wu, and Gan (2015), financial assets consist not only of risk-free assets including cash, bonds, demand deposits and fixed deposits, but also of risky assets including stocks, funds, financial derivatives, financial products, foreign exchange and gold.

The core explanatory variable concerned in this paper is "houseinvestment", that is, the impact of household housing investment on household stock investment. The value-added part of the housing investment is set as "houseprofit". This paper takes it as one of the control variables, and another corresponding variable, the value-added part after the stock investment "stockprofit" is also one of the control variables in this paper.

In reality, many factors have a bearing for the household participation in the stock market and stock holding ratio. Therefore, we refer to relevant literature and construct a series of other control variables: age and its square of the property owner that reflect life cycle effect (Wang &

Tian, 2012), gender of the property owner, marital status, education level (Russell, 2013), number of families, household income (Wang & Tian, 2012), financial literacy (Lusardi & Mitchell, 2007; Calvet et al., 2009; Hurd et al., 2011; Guan, Zhang, & Huang, 2013; Yin et al., 2014).

Among them, the definitions of financial literacy and risk attitude are particularly noteworthy.

Financial literacy corresponds to the number of interest rate questions, inflation questions and investment risk questions correctly answered by respondents (Agnew & Szykman, 2005). Namely, one correct answer means one point will be given.

Risk attitude corresponds to the question "if you have a sum of money, what kind of investment project would you like to choose?" In this regard, risk attitude is divided into five levels. According to Zhou (2015) setting, this paper redefines risk attitude variables. We classify 1 (projects with high risk and high return) and 2 (project with slightly high risk and slightly high return) as the same group that is referred to as risk-loving family, and the variable is set at 1; 3 (projects with average investment risk and average return) as risk neutral families, the variables are 0; 4 (projects with slightly lower risk and slightly lower return) and 5 (those who are unwilling to bear any risk) as risk averse families, the variables are - 1 (Zhou, 2015).

In Table 1, we give descriptive statistics of the variables used in this paper.

As table 1 shows, the stock market participation among residents is quite low. Further processing reveals that of the 13983 observation values, only 1699 households have stock accounts, accounting for only 12.15%, and the proportion of households' stockholding is also very low, with an average of 1.9%.

From the two variables, namely, stock profit and loss and housing profit and loss, it is clear that the average value of stock profit and loss is negative, and the average value of housing profit and loss is positive. Further analysis shows that in the observed samples, the proportion of households who have stock loss is 70.4%, while the proportion of investment stock profit is only 14.97%. By contrast, the proportion of investment property loss is only 5.43%, while the proportion of investment

HOUSING INVESTMENT, STOCK MARKET PARTICIPATION AND HOUSEHOLD PORTFOLIO CHOICE    10property profit is 71.24%. This data shows that most of the residents' stock investment is at a loss, and most of the investment in housing is at a profit. Because the yield record will affect the residents' expected yield on the two assets, it also foreshadows the measurement results.

Table 1

*Descriptive Statistics of Variables*

| Variables | description | Mean | Std. Error | Min | Max | N obs. |
|---|---|---|---|---|---|---|
| ifstock | Stock market participation | 0.122 | 0.327 | 0 | 1 | 13983 |
| capital | Household investment assets (million yuan) | 0.129 | 0.375 | 0.000 | 8.792 | 28414 |
| stock_ratio | Proportion of shares in investment assets | 0.019 | 0.074 | 0.0 | 0.5 | 13983 |
| house_ratio | Proportion of housing in investment assets | 0.544 | 0.418 | 0 | 1 | 13983 |
| houseinvestment | Housing investment cost (million yuan) | 0.169 | 0.367 | 0.0 | 7.2 | 13983 |
| housewealth | Housing wealth (million yuan) | 0.564 | 0.902 | 0.0 | 10.6 | 13983 |
| houseprofit | Housing investment profit (million yuan) | 0.395 | 0.772 | -4.94 | 10.200 | 13983 |
| stockprofit | Profit from stock investment (10000 yuan) | -0.323 | 5.793 | -350.00 | 170 | 13983 |
| stock_earning | Stock gains and losses | -0.554 | 0.740 | -1 | 1 | 1436 |
| house_earning | Property profit and loss | 0.658 | 0.578 | -1 | 1 | 13983 |
| education | Education (years) | 15.588 | 4.936 | 0 | 21 | 13983 |
| finance_edu | Financial education | 0.106 | 0.308 | 0 | 1 | 13978 |
| know_finance | Financial literacy | 0.539 | 0.630 | 0 | 2 | 13983 |
| risk_attitude | Risk attitude | -0.571 | 0.674 | -1 | 1 | 13859 |
| age | Age | 52.190 | 14.150 | 17 | 100 | 13983 |
| gender | Gender | 0.716 | 0.451 | 0 | 1 | 13983 |
| marriage | Marriage | 0.877 | 0.328 | 0 | 1 | 13983 |
| number | Household size | 3.344 | 1.418 | 1 | 14 | 13983 |
| familyincome | Household income (10000 yuan) | 7.974 | 11.916 | 0.00 | 262.53 | 13983 |
| familywealth | Household wealth (million yuan) | 1.119 | 1.608 | 0.000 | 19.963 | 13983 |

The average value of financial education is very low. Further analysis shows that only 10.58% of the respondents have received economic or financial courses. Compared with the variable of financial literacy, 53.55% of the residents answered the questions of interest rate, inflation and investment risk completely wrong, 39.03% of the respondents could answer one of the questions correctly, 7.42% of the residents could answer two of the questions and three of the questions correctly. No household got all the questions correctly. This shows that Chinese families lack the basic understanding of finance and financial market, which also foreshadows the measurement results.

The mean value of risk attitude is negative. Further analysis shows that 67.62% of the samples are risk averse families.

The average age is 52, which is 4 years older than the average age of 48 in CHFS2011. This paper preliminarily estimates that the majority of property owners between 2011 and 2013 are middle-aged over-48s.

The average family member number is 3, which is in line with the family planning policy implemented in China.

The average annual household income is 80,000 yuan, and the average value plus a standard deviation is 200,000 yuan, which indicates that the household income in the sample, namely, the annual income of property-owning urban residents is mostly within 200,000 yuan.

## Empirical Specifications

**Mathematical Model**

The impact of housing investment on household stock market participation is examined first, followed by analyzing the impact of housing investment on the proportion of household stock assets in total investment assets.

In a bid to study the impact of housing investment on stock market participation, we need to use Probit regression. Suppose for the household i, its decision to participate in the stock market $Y_i$ is either 0 or 1. And the value is jointly decided by the latent variable $Y_i^*$ ($Y_i^* > 0$, $Y_i = 1$; $Y_i^* \leq 0$, $Y_i = 0$) whose value is determined by $houseinvestment_i$, other control variables $X_i$, and random variables $\varepsilon_i$ that describe invisible household traits:

$$Y_i = 1(Y_i^* > 0) \tag{1}$$

$$Y_i^* = \alpha_1 \cdot houseinvestment_i + \sum_f \beta_f X_i + \varepsilon_i \tag{2}$$

Here, $\varepsilon_i$ obeys the Gaussian distribution, $\alpha_1$、$\beta_f$ are the estimated parameters. For (2), we can use maximum likelihood estimation to estimate these parameters.

As the proportion $Z_i$ of stock assets in investment assets is truncated, ranging between [0,1], Tobit regression is used so as to test the impact of housing investment. The proportion of risky assets held by the household $i$ to total non-housing assets is recorded as $Z_i$, then it is determined by $houseinvestment_i$, other control variables $X_i$, and invisible family traits $\eta_i$.

$$Z_i = max\{\alpha_2 \cdot houseinvestment_i + \sum_k \beta_k X_i + \eta_i, 0\} \quad （3） \quad (3)$$

Here, $\alpha_2$ and $\beta_k$ are the parameters to be estimated. Likewise, we can also use the maximum likelihood estimation to estimate (3).

**Benchmark Regression**

This paper applies the mathematical model above to the explanatory variable of housing investment and the control variables of housing investment profit, stock investment profit, education, financial literacy, age, square of age, gender, marriage, number of families, household income and household wealth, and obtains significant regression results.

As Zhou (2015) pointed out that after grouping families according to risk attitude, the effect of financial education on the financial markets participation is asymmetric. Therefore, the whole sample is divided into three categories to improve the model precision—risk aversion, risk neutrality and risk loving — to explore the link between household housing investment and household stockholding.

Therefore, in terms of research methods, this paper conducts another Probit and Tobit regression for each kind of households. However, the Probit regression results of stock market participation in risk averse families are not convergent, so this paper uses OLS regression instead.

**Overcoming Endogenous Problems**

**Heckman Model checking.** Considering the self-selection of samples and the fact that stock market value belongs to censored data, there may be endogenous problems that should be tested in this paper with the two-step estimation method of Heckman model.

Specifically, the first step is to make a Probit estimate of whether or not to hold a stock account; the second step is to take the Inverse Mills Ratio obtained from the Probit estimate in the first step as both a correction parameter in the second stage and an additional variable of the equation estimate to correct the selectivity bias, and then make a Heckman model estimate of the stock value.

**Regression based on subsample.** Granted, after the Heckman model test, the endogenous problem still cannot be completely eliminated. For example, some residents only buy houses for needs, not just for investment.

In response to this problem, samples of properties purchased after 1998 are employed to apply the Heckman model. In the history of China's housing market, the "National Conference on Urban Housing System Reform and Housing Construction" in 1998 serves as an important watershed event. Afterwards, the state's housing policy no longer emphasizes "welfare". Therefore, the changing pattern of the housing price switched from very gentle to a rapid pattern, and the residents began to buy houses for investment.

## Results and Discussion

**Benchmark Regression**

In Table 2, we give the benchmark regression results of formula (1) - (3). Among them, columns (I), (III) and (V) respectively show the Probit regression results of the impact of housing on household stock market participation; column (VII) shows the OLS regression results, while columns (II), (IV), (VI) and (VIII) show the regression results of the impact of housing on household stockholding. In order to describe the impact of housing more clearly, we give marginal effects in all the tables in Table 2.

Firstly, the impact of housing on household stock market participation. In column (I), after controlling all kinds of control variables, the marginal effect of housing investment cost is significantly positive at the level of 1%, which indicates that housing will improve the probability of Chinese household stock market participation. The marginal effect of 0.045 suggests that the probability of household holding stock will increase by 4.5% when the housing investment increases a sample standard deviation, namely 367,000 yuan. In our sample, the probability of household stockholding is only 12.2%. We can see that the contribution of housing investment to household stock market participation is significant.

Table 2

*Results of Benchmark Regression*

| | (I) | (II) | (III) | (IV) | (V) | (VI) | (VII) | (VIII) |
|---|---|---|---|---|---|---|---|---|
| Response variable | | | Risk preference | | Risk neutral | | Risk averse | |
| | participation | Share of stock | participation | Share of stock | participation | Share of stock | participation | Share of stock |
| Housing investment | 0.045*** (0.011) | -0.038*** (0.013) | 0.037 (0.041) | -0.052** (0.022) | 0.054** (0.025) | -0.047*** (0.018) | 0.048*** (0.017) | -0.029 (0.024) |
| Square of housing investment | -0.010*** (0.003) | \ | -0.007 (0.009) | \ | -0.015** (0.007) | \ | -0.006 (0.007) | \ |
| Profit from housing investment | 0.008** (0.003) | 0.013** (0.007) | 0.039*** (0.015) | 0.032** (0.014) | 0.019** (0.009) | 0.010 (0.011) | 0.004 (0.005) | 0.019* (0.01) |
| Profit from stock Investment | -0.005*** (0.000) | -0.007*** (0.001) | -0.002** (0.001) | -0.002*** (0.001) | -0.009*** (0.001) | -0.011*** (0.001) | -0.018*** (0.001) | -0.017*** (0.001) |
| Education | -0.002*** (0.001) | -0.004*** (0.001) | -0.001 (0.003) | -0.001 (0.003) | -0.005*** (0.002) | -0.005** (0.002) | -0.002*** (0.001) | -0.004** (0.002) |
| Financial literacy | 0.064*** (0.004) | 0.115*** (0.008) | 0.100*** (0.016) | 0.08*** (0.017) | 0.074*** (0.010) | 0.078*** (0.013) | 0.045*** (0.005) | 0.113*** (0.013) |
| Age | 0.008*** (0.001) | 0.015*** (0.003) | 0.019*** (0.005) | 0.022*** (0.006) | 0.009*** (0.003) | 0.009** (0.004) | 0.004*** (0.001) | 0.014*** (0.004) |
| Square of age | -0.000*** 0.000 | 0.000*** 0 | -0.000*** (0.000) | 0.000*** (0.000) | -0.000** (0.000) | 0.000* (0.000) | 0.000*** (0.000) | 0.000*** (0.000) |
| Gender | -0.002 (0.006) | -0.009 (0.012) | 0.001 (0.026) | 0.005 (0.025) | -0.042*** (0.014) | -0.058*** (0.019) | -0.001 (0.006) | -0.013 (0.017) |
| Marriage | 0.023** (0.009) | 0.045** (0.018) | 0.085** (0.037) | 0.054 (0.037) | 0.035 (0.024) | 0.034 (0.031) | 0.009 (0.009) | 0.061** (0.028) |
| Household size | -0.018*** (0.002) | -0.034*** (0.005) | -0.052 (0.009) | -0.042*** (0.010) | -0.027*** (0.005) | -0.031*** (0.008) | -0.010*** | -0.034*** (0.007) |
| Household income | 0.002*** (0.000) | 0.002*** 0 | 0.002* (0.001) | 0.001 (0.001) | 0.001** (0.000) | 0.001 (0.001) | 0.003*** (0.000) | 0.004*** (0.001) |
| Household wealth | 0.020*** (0.002) | 0.039*** (0.004) | 0.022*** (0.007) | 0.026*** (0.007) | 0.028*** (0.004) | 0.037*** (0.006) | 0.021*** (0.003) | 0.032*** (0.006) |
| N obs. | 13983 | 13983 | 1453 | 1453 | 3034 | 3034 | 9372 | 9372 |

*Note*: (1) *, **, *** are significant at the level of 10%, 5% and 1% respectively;

(2) the number in brackets is the standard error of marginal effect.

Nevertheless, it is noteworthy that in column (II), the marginal effect of housing on the proportion of household stock holdings is significantly negative at the level of 1%, which indicates that housing investment will reduce the proportion of Chinese household stockholding. The marginal effect of 0.038 tells that the proportion of household stock holdings will decrease by 3.8% if the housing investment increases by a sample standard deviation, that is, 367,000 yuan. In our sample, the average share holding ratio of households is only 1.9%, so the inhibition of housing investment on the household share holding ratio is significant, too.

In the sub samples with the same risk preference, that is, the marginal effect of household housing investment on the proportion of household stockholding in columns (IV), (VI) and (VIII) is negative, and in columns (IV) and (VI), it is significantly negative at the level of 1%, which proves the crowding out effect of housing investment on stock investment. Moreover, it can be concluded that the substitution effect of housing investment and stock investment is very strong when the risk-loving household makes the investment decision.

Hence, what causes the probability of household stock market participation to rise with housing investment, while the proportion of household stock holdings to decline? It is easy to associate that in the descriptive statistics of variables, the average value of stock market profit and loss is -0.554, that is to say, among the families participating in stock market investment, 70.4% of the household stock accounts lost money, only 14.97% of the household stock accounts gained profit. This data is quite striking, even if the stock price is regarded as a random walk, this profit and loss data is anomalous. Correspondingly, the average profit and loss of housing is 0.658. That is, 71.24% of the families participating in the housing investment made profit, and only 5.43% of the families lost money.

Moreover, in (I), (II), (III), (IV), (V), (VI), (VII) and (VIII), the marginal effect of stock investment profit on household stock market participation and stock holding ratio is significantly negative. In (I) and (II), they are -0.005 and -0.007, respectively. Obviously, this data cannot be interpreted as a standard deviation increase of stock investment profit, that is, an increase of 57,930 yuan, a decrease of 0.5% in the probability of household stock market participation and a decrease of 0.7% in the proportion of household stock holdings. Therefore, given the average value of stock investment profit is -0.323 means the loss is 3,230 yuan and the probability of household stock loss in the sample is as high as 70.4%, this data can be interpreted as that, in the families participating in the stock market, the larger the share holding ratio is, the more the loss will be. Combined with the contribution of housing investment to household stock market participation, it can be inferred that housing investment drives stock market investment and prompts more residents to hold stocks, but their attempts to invest in stocks often fail. Even worse, the more they invest, the more they lose, so when they have new income, they will not invest or invest less in the stock market.

By contrast, the continuously rising house prices in the past decade has made residents feel that housing is almost risk-free, the residents' awareness of risk prevention is reduced, and the overall degree of housing risk preference tends to be positive. Therefore, the historical yield of two kinds of investment assets determines their expected yield.

Given that the impact of financial literacy on household stock market participation and share holding ratio is significantly positive, and the marginal effect is 0.064 and 0.115 respectively, we can see that financial literacy plays a very crucial role in promoting household's stock market participation and share holding ratio. Integrated with the lack of household financial computing power reflected in descriptive statistics of financial literacy, it can be concluded that the popularization of financial literacy of Chinese residents fails to meet the growing investment demand. After continuous investment failures in stock market, their expected return on stocks will continue to decline, and their new investment funds will only flow less into the stock market, and more into the booming housing sector.

In addition, the regression results also tells the impact of other factors on the household stock market participation: first, the profit from housing investment will significantly push up the probability of household participation in the stock market, and its marginal effect is 0.008, which shows that the probability of household participation in the stock market can be increased by 0.8% by adding a standard deviation (i.e. RMB 7.72 million) to the profit from housing investment (sample participation ratio is 12.2%). Moreover, the marginal effect of housing investment profit on the proportion of household stockholding is 0.013, which means that a standard deviation (RMB 7.72 million) increase of housing investment profit can increase the proportion of household stockholding by 1.3%.

Education has a significant negative impact on the probability of household stock market participation and the proportion of stock holdings, which confirms Wu Weixing's conclusion that the lack of education will lead to overconfidence of consumers that leads to the occurrence of transactions that should not have happened (Wu, Wang, & Liang, 2006). In the sample used in this paper, due to a general lack of financial literacy, the highly educated residents are aware of their limited financial literacy and have no blind confidence, so their stockholding is reduced.

Age has a significant positive impact on the probability of household stock market participation and the proportion of stockholding, which may indicate that the middle-aged and elderly people are willing to try new ways of investment when their wealth is gradually accumulating.

Marriage has a significant positive effect on the probability of household stock market participation and the proportion of stock holding, and the marginal effect is very large, which may be associated with the stable household structure that gives the confidence of the property owner to invest in risky assets. Nonetheless, the number of family member has a significant negative impact on the probability of household stock market participation and the proportion of stockholding, which also indicates that many family members will hinder household stock investment, probably because the large number of families are facing greater financial pressure.

The regression results of residents with different risk attitudes also reflect these phenomena, which will not be discussed here.

**Tackling Endogenous Problems**

**Heckman model.** Although Probit and Tobit regression have shown that higher housing investment is crowding out household stock investment, the above conclusion may be disturbed by some endogenous problems. In order to eliminate the interference, we tested it with Heckman model. In this model, the third column also introduces the cross term of housing investment cost and financial literacy to study the joint effect of the two on the proportion of household stock investment. In Table 3, we show the results of these regressions.

Among them, columns (II) and (IV) are the selection equations, namely, apply Probit model in the first step to examine if they hold stock accounts; columns (I) and (III) are the Heckman model estimation results that take the inverse mills ratio obtained from the first step Probit estimation as another variable of the equation. Similarly, all we give in tables in Table 3 are marginal effects.

Table 3

Test of Heckman Model

|  | (I) Heckman | (II) Choice model | (III) Heckman | (IV) Choice model |
|---|---|---|---|---|
| Housing investment | 0.086*** (0.007) | 0.045*** (0.011) | -0.08*** (0.011) | 0.059*** (0.014) |
| Housing investment squared |  | -0.010*** (0.003) |  | -0.01*** (0.003) |
| Financial literacy | 0.004 (0.008) | 0.064*** (0.004) | 0.007 (0.009) | 0.069*** (0.004) |
| Housing investment *Financial literacy |  |  | -0.006 (0.008) | -0.014* (0.008) |
| Profit from housing investment | -0.007* (0.004) | 0.008** (0.003) | -0.007* (0.004) | 0.009*** (0.003) |
| Profit from stock investment | -0*** (0.000) | -0.005*** (0.000) | -0*** (0.000) | -0*** (0.000) |
| Education | 0.001 (0.001) | -0.002*** (0.001) | 0.001 (0.001) | -0.002*** (0.001) |
| Age | 0.003* (0.002) | 0.008*** (0.001) | 0.003* (0.002) | 0.007*** (0.001) |
| Square of age | 0.000 (0.000) | -0.000*** (0.000) | -0 (0.000) | -0*** (0.000) |
| Gender | 0.010 (0.007) | -0.002 (0.006) | 0.010 (0.007) | -0.003 (0.006) |
| Marriage | -0.012 (0.012) | 0.023** (0.009) | -0.012 (0.012) | 0.019** (0.009) |
| Household size | -0.004 (0.003) | -0.018*** (0.002) | -0.005 (0.003) | -0.014*** (0.002) |
| Household income | 0.000 (0.000) | 0.002*** (0.000) | 0.000 (0.000) | 0.002*** (0.000) |
| Household wealth | 0.009*** (0.002) | 0.020*** (0.002) | 0.009*** (0.002) | 0.02*** (0.002) |
| Inverse Mills Ratio | 0.027 (0.020) |  | 0.030 (0.020) |  |
| rho | 0.194 |  | 0.213 |  |
| N obs. | 13983 | 13983 | 13983 | 13983 |

Note: (1) *, * *, * * * are significant at the level of 10%, 5% and 1% respectively;

(2) the number in brackets is the standard error of marginal effect.

As column (I) tells, the marginal effect of housing investment is significantly negative at the level of 1%, showing that the increase of housing investment will still significantly reduce the proportion of household stock investment after reducing the interference of endogenous problems. From the numerical point of view, the marginal effect of housing investment is -0.086, which is significantly greater than the marginal effect of Probit regression, more than twice. According to this value, when the housing investment jumps a sample standard deviation, the household stockholding drops by 8.6%, which also has a strong economic implication.

It is noticeable that the marginal effect of housing investment profit changes from significantly positive to significantly negative from column (I). This shows that after removing sample selection's endogeneity, the wealth effect of housing investment profit disappears, and its increase will significantly reduce the proportion of household stock investment. From the numerical point of view, the marginal effect of housing investment profit is -0.007. That is, if the cost of housing investment increases by a sample standard deviation, the proportion of shares held by households will fall by 0.7%. Evidently, this value also has a strong economic significance.

This paper speculates that this is because of the fact that many families do not have a stock account, but there are housing investment costs and housing investment profits. When this paper only uses Tobit regression, this part of families have no impact on Tobit regression. However, when this paper uses Heckman model to combine the predicted probability of each household's participation in the stock market into an additional explanatory variable, together with other variables to correct the self selection problem in the household equity investment proportion model, these account-lacking families has a great impact on the regression. This demonstrates that after removing the endogeneity of sample selection, the crowding out effect of housing investment on household stock investment is greater. Because many families have a strong affinity for housing investment, they only invest in housing and have no stock investment account at all!

As shown the third column, after introducing the cross item of housing investment cost and financial literacy, the marginal effect of housing investment cost and housing investment profit are still significantly negative, and the value of marginal effect is similar to that of the first column. This result reconfirms the conclusion that housing investment squeezes out the proportion of household stock investment. The marginal effect of the cross term between housing investment and financial literacy is negative, which shows that even families with financial literacy still prefer housing in their investment decision-making, verifying the conclusion drawn that the popularity of residents' financial literacy cannot keep up with the growing investment demand.

**Subsample.** In order to further eliminate the interference of endogenous problems, we focus on the subsamples of properties purchased after 1998. Table 4 shows the results of Heckman model test based on the above sub sample.

Table 4

Test of Heckman Model Based on the Samples of Houses Purchased After 1998

|  | (I) Heckman | (II) Choice model | (III) Heckman | (IV) Choice model |
|---|---|---|---|---|
| Housing investment | -0.074*** (0.008) | 0.045*** (0.012) | -0.054*** (0.013) | 0.062*** (0.016) |
| Housing investment squared |  | -0.009** (0.003) |  | -0.009*** (0.003) |
| Financial knowledge | 0.027** (0.013) | 0.069*** (0.005) | 0.037** (0.015) | 0.073*** (0.005) |
| Housing investment *Financial literacy |  |  | -0.02** (0.01) | -0.016* (0.01) |
| Profit from housing investment | -0.013** (0.005) | 0.006 (0.004) | -0.013** (0.005) | 0.006 (0.004) |
| Profit from stock investment | -0*** (0.000) | 0*** (0.000) | -0*** (0.000) | -0*** (0.000) |
| education | -0.001 (0.001) | -0.003*** (0.001) | -0.001 (0.001) | -0.003*** (0.001) |
| Age | 0.008*** (0.002) | 0.006*** (0.002) | 0.009*** (0.002) | 0.006*** (0.002) |
| Square of age | -0*** (0.000) | -0*** (0.000) | -0*** (0.000) | -0*** (0.000) |
| Gender | 0.018* (0.009) | 0.005 (0.007) | 0.019** (0.01) | 0.005 (0.007) |
| Marriage | -0.015 (0.016) | 0.019* (0.012) | -0.014 (0.016) | 0.019* (0.012) |
| Household size | -0.003 (0.005) | -0.016*** (0.003) | -0.004 (0.005) | -0.016*** (0.003) |
| Household income | 0.000 (0.000) | 0.001*** (0.000) | 0.001* (0.000) | 0.001*** (0.000) |
| Household wealth | 0.017*** (0.004) | 0.022*** (0.002) | 0.017*** (0.004) | 0.022*** (0.002) |
| Inverse Mills Ratio | 0..104*** (0.038) |  | 0.115*** (0.040) |  |
| rho | 0.642 |  | 0..687 |  |
| N obs. | 9698 | 9698 | 9698 | 9698 |

Note: (1) *, * *, * * * are significant at the level of 10%, 5% and 1% respectively;

(2) the number in brackets is the standard error of marginal effect.

As can be seen from the results of each column in Table 4, the significance of the model is greatly improved for this sub sample. In column (I) and column (III), the marginal effect of housing investment cost is -0.074 and -0.054 respectively, and the absolute value is smaller than the result of Heckman model test with the whole sample. It can be speculated that, because the families that bought houses before 1998 gained great value-added, their preference for housing investment is greater than that of the families who bought houses after 1998. Consequently, the proportion of housing investment squeezed out more household stockholding for the families that bought houses before 1998.

With regards to housing investment profit, its marginal effect is significantly negative, and its absolute value is as high as 0.013, which further shows that after removing the endogenous problem, the "crowding out effect" of housing wealth will become more prominent.

It is worth noting that for this subsample, the marginal effect of the cross term of housing investment cost and financial literacy is significantly negative at the level of 5% in column (III), and the absolute value of the marginal effect is 0.02, only 0.034 less than the absolute value of the marginal effect of housing investment cost! This highlights that this index cannot be ignored in economic sense, and further underscores that even families with financial expertise still prefer housing in investment decisions.

In conclusion, the inevitable existence of endogenous problems makes us underestimate the impact of housing investment on household stock investment. After controlling the endogenous problem, the "crowding out effect" has become more significant, and the underlying cause inferred in this paper has been confirmed more clearly.

## Conclusions

In this paper, CHFS data is used to examine the implication of housing investment on household financial market participation and portfolio choice. We find that higher housing investment will significantly boost household stock market participation, but will drop the proportion of stockholding. This conclusion still holds when the endogeneity problem outstripped by risk attitude classification, Heckman model test and subsample regression.

It is worth noting that although housing investment will promote the likelihood of stock market participation, we cannot be blindly optimistic that the development of housing will lead to a more prosperous stock market. Because the yield record of two kinds of investment assets suggests that the housing investment drives the stock market investment and stimulates more residents to hold stocks, but their attempts to invest in stocks often fail. Even worse, the more they invest, the more they can lose. So, they will not invest or invest small in the stock market with new investment funds.

Combined with the obvious lack of household financial literacy across China, it can be concluded that the popularization of financial literacy of Chinese residents fails to keep up with the growing investment demand. After continuous setbacks in stock market, their expected yield on stocks will continue to decline, and their new investment funds will only flow less into the stock market, and more into the booming housing sector. Therefore, we need to popularize financial education to the general public and build their financial literacy, so that their investment in stocks can be restored to health, and stock investment can play in a "level playing ground" with housing investment.

HOUSING INVESTMENT, STOCK MARKET PARTICIPATION AND HOUSEHOLD PORTFOLIO CHOICE                    24